\documentclass[aps,prd,twocolumn,showpacs,nofootinbib]{revtex4-1}

\usepackage{graphicx}
\usepackage{amssymb}
\usepackage{color}

\usepackage[colorlinks=true,linkcolor=blue,citecolor=blue, urlcolor=blue]{hyperref}

\begin{document}

\title{The $\Upsilon(1S)$ leptonic decay using the principle of maximum conformality}

\author{Xu-Dong Huang$^1$}
\email{hxud@cqu.edu.cn}
\author{Xing-Gang Wu$^1$}
\email{wuxg@cqu.edu.cn}
\author{Jun Zeng$^1$}
\email{zengj@cqu.edu.cn}
\author{Qing Yu$^1$}
\email{yuq@cqu.edu.cn}
\author{Jian-Ming Shen$^2$}
\email{shenjm@hnu.edu.cn}

\affiliation{$^1$ Department of Physics, Chongqing University, Chongqing 401331, People's Republic of China}
\affiliation{$^2$ School of Physics and Electronics, Hunan University, Changsha 410082, People's Republic of China}

\date{\today}

\begin{abstract}

In the paper, we study the $\Upsilon(1S)$ leptonic decay width $\Gamma(\Upsilon(1S)\to \ell^+\ell^-)$ by using the principle of maximum conformality (PMC) scale-setting approach. The PMC adopts the renormalization group equation to set the correct momentum flow of the process, whose value is independent to the choice of the renormalization scale and its prediction thus avoids the conventional renormalization scale ambiguities. Using the known next-to-next-to-next-to-leading order perturbative series together with the PMC single scale-setting approach, we do obtain a renormalization scale independent decay width, $\Gamma_{\Upsilon(1S) \to e^+ e^-} = 1.262^{+0.195}_{-0.175}$ keV, where the error is squared average of those from $\alpha_s(M_{Z})=0.1181\pm0.0011$, $m_b=4.93\pm0.03$ GeV and the choices of factorization scales within $\pm 10\%$ of their central values. To compare with the result under conventional scale-setting approach, this decay width agrees with the experimental value within errors, indicating the importance of a proper scale-setting approach.

\end{abstract}

\maketitle

Since the $b$-quark mass is much larger than the QCD asymptotic scale, $m_b>>\Lambda_{\rm QCD}$, the leptonic decay of the heavy quarkonium $\Upsilon(1S)$ is one of the important channel for testing the non-relativistic QCD theories. At present, the decay width $\Gamma(\Upsilon(1S)\to e^+ e^-)$ has been calculated up to next-to-next-to-next-to-leading order (N$^3$LO) level~\cite{Pineda:1996uk, Pineda:2001et, Beneke:1999fe, Pineda:2006ri, Beneke:2014qea, Marquard:2014pea, Beneke:2005hg, Penin:2005eu, Beneke:2007gj, Beneke:2007pj, Beneke:2008cr, Mutuk:2018xay}. At the N$^3$LO level, the conventional renormalization scale uncertainty is still very large, which is usually estimated by varying the renormalization scale ($\mu_r$) within the assumed range of $[3, 10]$ GeV. However, at this perturbative order, the predicted decay width is still lower than the PDG averaged experimental value~\cite{Beneke:2014qea}, i.e. $\Gamma_{\Upsilon(1S)\to e^+ e^-}|_{\rm Exp.} = 1.340(18)$~keV~\cite{Tanabashi:2018oca}. It has been pointed out that the conventional scale-setting approach, in which the renormalization scale is guessed and usually chosen as the one to eliminate the large logs, will meet serious theoretical problems due to the mismatching of $\alpha_s$ and the coefficients at each perturbative order, and its accuracy depends heavily on the how many terms of the pQCD series are known and the convergence of the pQCD series~\cite{Wu:2013ei}. It is thus important to adopt a proper scale-setting approach so as to achieve a more accurate fixed-order pQCD prediction.

In year 2015, the authors of Ref.\cite{Shen:2015cta} used the principle of maximum conformality (PMC)~\cite{Brodsky:2011ta, Brodsky:2011ig, Mojaza:2012mf, Brodsky:2013vpa} to eliminate such scale ambiguity and predicted, $\Gamma_{\Upsilon(1S) \to e^+ e^-} \sim 1.27$~keV. This value agrees with experimental value within errors by further considering the factorization scale uncertainty. However, the analysis there was done by using the PMC multi-scale approach (PMC-m)~\cite{Mojaza:2012mf, Brodsky:2013vpa}, in which the PMC scales at each order are different and are of perturbative nature whose values for higher-order terms are of less accuracy due to more of its perturative terms are unknown, leading to a somewhat larger residual scale dependence. More explicitly, for a N$^3$LO-level pQCD series of $\Gamma_{\Upsilon(1S) \to e^+ e^-}$, the PMC-m approach shows that there are three PMC scales for its LO, NLO and NNLO terms accordingly~\cite{Shen:2015cta}: the LO PMC scale is at the N$^2$LL accuracy, the NLO PMC scale is at the NLL accuracy and the NNLO PMC scale is at the LL accuracy, respectively.

Recently, a single-scale PMC scale-setting approach (PMC-s) has been suggested by Ref.\cite{Shen:2017pdu}, which fixes the scale by using all the $\beta$-terms of the process as a whole and can achieve a scale independent and scheme independent prediction at any fixed order, satisfying the renormalization group invariance~\cite{Wu:2018cmb}. Since such scale is determined by using the renormalization group equation, it determines an effective value of the strong coupling constant $\alpha_s(Q_*)$, whose argument $Q_*$ corresponds to an overall effective momentum flow of the process. In this paper, as an attempt, we adopt the PMC-s approach with the purpose of achieving a more accurate pQCD prediction free of renormalization scale error on the $\Upsilon(1S)$ leptonic decay width.

Up to N$^3$LO-level, the decay width $\Gamma_{\Upsilon(1S)\to \ell^+\ell^-}$ can be written in the following form by using the degeneracy relations among different orders~\cite{Mojaza:2012mf, Brodsky:2013vpa}
\begin{eqnarray}
\Gamma_3 &=& r_{1,0}a_s^3(\mu_{r}) + (r_{2,0}+3\beta_{0}r_{2,1})a_{s}^{4}(\mu_{r})  \nonumber\\ &&+(r_{3,0}+3\beta_{1}r_{2,1}+ 4\beta_{0}r_{3,1}+ 6\beta_{0}^{2}r_{3,2})a_{s}^{5}(\mu_{r})\nonumber\\
&& +(r_{4,0}+3\beta_{2}r_{2,1}+ 4\beta_{1}r_{3,1} +5\beta_{0}r_{4,1} + \nonumber\\
&& \frac{27}{2}\beta_{1}\beta_{0}r_{3,2} +10\beta_{0}^{2}r_{4,2}+10\beta_{0}^{3}r_{4,3}) a_{s}^{6}(\mu_{r}), ~\label{rij}
\end{eqnarray}
where $a_s=\alpha_s/4\pi$, and the coefficients $r_{i,j}$ can be derived from Refs.\cite{Marquard:2014pea, Beneke:2005hg, Penin:2005eu, Beneke:2007gj, Beneke:2007pj, Beneke:2008cr}, whose explicit expressions have been given in the Appendix of Ref.\cite{Shen:2015cta}. The conformal coefficients $r_{i,0}=\hat{r}_{i,0}$ are independent of the initial choice of renormalization scale $\mu_r$, and the non-conformal coefficients $r_{i,j}$ $(j\neq0)$ are functions of $\mu_r$, which can be written as
\begin{eqnarray}
r_{i,j}=\sum^j_{k=0}C^k_j{\hat r}_{i-k,j-k}{\rm ln}^k(\mu_r^2/m_b^2),~\label{rijrelation}
\end{eqnarray}
where ${\hat r}_{i,j}=r_{i,j}|_{\mu_r=m_b}$, $C^k_j$ is defined as $j!/(k!(j-k)!)$, and $i,j,k$ are the polynomial coefficients. By substituting Eq.~(\ref{rijrelation}) into Eq.~(\ref{rij}), the decay width $\Gamma_{\Upsilon(1S)\to \ell^+\ell^-}$ can be written as
\begin{widetext}
\begin{eqnarray}
\Gamma_3 &=&{\hat r}_{1,0}a_s^3(\mu_{r}) + [{\hat r}_{2,0}+3\beta_{0}({\hat r}_{2,1}+{\hat r}_{1,0} \ln\frac{\mu_r^2}{m_b^2})]a_{s}^{4}(\mu_{r}) +[{\hat r}_{3,0}+3\beta_{1}({\hat r}_{2,1}+{\hat r}_{1,0} \ln\frac{\mu_r^2}{m_b^2})+ 4\beta_{0}({\hat r}_{3,1}\nonumber\\
&&+{\hat r}_{2,0} \ln\frac{\mu_r^2}{m_b^2})+ 6\beta_{0}^{2}({\hat r}_{3,2}+2{\hat r}_{2,1} \ln\frac{\mu_r^2}{m_b^2}+{\hat r}_{1,0} \ln^2\frac{\mu_r^2}{m_b^2})]a_{s}^{5}(\mu_{r})+[{\hat r}_{4,0}+3\beta_{2}({\hat r}_{2,1}+{\hat r}_{1,0} \ln\frac{\mu_r^2}{m_b^2})\nonumber\\
&& + 4\beta_{1}({\hat r}_{3,1}+{\hat r}_{2,0} \ln\frac{\mu_r^2}{m_b^2}) +5\beta_{0}({\hat r}_{4,1}+{\hat r}_{3,0} \ln\frac{\mu_r^2}{m_b^2}) +\frac{27}{2}\beta_{1}\beta_{0}({\hat r}_{3,2}+2{\hat r}_{2,1} \ln\frac{\mu_r^2}{m_b^2}+{\hat r}_{1,0} \ln^2\frac{\mu_r^2}{m_b^2}) \nonumber\\
&&+10\beta_{0}^{2}({\hat r}_{4,2}+2{\hat r}_{3,1} \ln\frac{\mu_r^2}{m_b^2}+{\hat r}_{2,0} \ln^2\frac{\mu_r^2}{m_b^2})+10\beta_{0}^{3}({\hat r}_{4,3}+3{\hat r}_{3,2} \ln\frac{\mu_r^2}{m_b^2}+3{\hat r}_{2,1} \ln^2\frac{\mu_r^2}{m_b^2}+{\hat r}_{1,0} \ln^3\frac{\mu_r^2}{m_b^2})] a_{s}^{6}(\mu_{r}). ~\label{rqij}
\end{eqnarray}
\end{widetext}

Applying the standard PMC-s procedures~\cite{Shen:2017pdu}, all the non-conformal terms should be resummed into the running coupling. The N$^3$LO-level leptonic decay width $\Gamma_3$ changes to the following conformal series,
\begin{eqnarray}
\Gamma_3|_{\rm PMC-s} =&& {\hat r}_{1,0}a_s^3(Q_*) + {\hat r}_{2,0}a_s^4(Q_*)+{\hat r}_{3,0}a_s^5(Q_*)\nonumber \\
&&+ {\hat r}_{4,0}a_s^6(Q_*),~\label{conformal}
\end{eqnarray}
where $Q_*$ is the PMC scale that determines the effective momentum flow and hence the effective running coupling $\alpha_s(Q_*)$ of the process. More explicitly, the PMC scale $Q_*$ is obtained by first shifting the scale $\mu_r$ in $a_s$ to $Q_*$ in Eq.(\ref{rqij}) by using scale displacement relation of the strong coupling constant, i.e.
\begin{displaymath}
a^k_s(\mu_r) = a^k_s(Q_*) - k \beta_0 \ln \frac{\mu_r^2}{Q_*^2} a^{k+1}_s(Q_*)+\mathcal{O}[a^{k+2}_s(Q_*)].
\end{displaymath}
Then the PMC scale $Q_*$ is obtained by requiring all the non-conformal terms vanish, e.g.
\begin{widetext}
\begin{eqnarray}
0 &=& 3\beta_{0}[{\hat r}_{2,1}+{\hat r}_{1,0}(\ln\frac{\mu_r^2}{m_b^2}+ \ln\frac{Q_*^2}{\mu_r^2})]a_{s}^{4}(Q_*) +\{3\beta_{1}[{\hat r}_{2,1}+{\hat r}_{1,0} (\ln\frac{\mu_r^2}{m_b^2}+ \ln\frac{Q_*^2}{\mu_r^2})]+ 4\beta_{0}[{\hat r}_{3,1}+{\hat r}_{2,0} (\ln\frac{\mu_r^2}{m_b^2}+ \ln\frac{Q_*^2}{\mu_r^2})]\nonumber\\
&& + 6\beta_{0}^{2}[{\hat r}_{3,2}+2{\hat r}_{2,1} (\ln\frac{\mu_r^2}{m_b^2}+ \ln\frac{Q_*^2}{\mu_r^2})+{\hat r}_{1,0} (\ln\frac{\mu_r^2}{m_b^2}+ \ln\frac{Q_*^2}{\mu_r^2})^2]\}a_{s}^{5}(Q_*)+\{3\beta_{2}[{\hat r}_{2,1}+{\hat r}_{1,0} (\ln\frac{\mu_r^2}{m_b^2}+ \ln\frac{Q_*^2}{\mu_r^2})] \nonumber\\
&& + 4\beta_{1}[{\hat r}_{3,1}+{\hat r}_{2,0} (\ln\frac{\mu_r^2}{m_b^2}+ \ln\frac{Q_*^2}{\mu_r^2})] +5\beta_{0}[{\hat r}_{4,1}+{\hat r}_{3,0} (\ln\frac{\mu_r^2}{m_b^2}+ \ln\frac{Q_*^2}{\mu_r^2})] +\frac{27}{2}\beta_{1}\beta_{0}[{\hat r}_{3,2}+2{\hat r}_{2,1} (\ln\frac{\mu_r^2}{m_b^2}+ \ln\frac{Q_*^2}{\mu_r^2})\nonumber\\
&& +{\hat r}_{1,0} (\ln\frac{\mu_r^2}{m_b^2}+\ln\frac{Q_*^2}{\mu_r^2})^2]+10\beta_{0}^{2}[{\hat r}_{4,2}+2{\hat r}_{3,1} (\ln\frac{\mu_r^2}{m_b^2}+ \ln\frac{Q_*^2}{\mu_r^2})+{\hat r}_{2,0} (\ln\frac{\mu_r^2}{m_b^2}+\ln\frac{Q_*^2}{\mu_r^2})^2]+10\beta_{0}^{3}[{\hat r}_{4,3}\nonumber\\
&& +3{\hat r}_{3,2} (\ln\frac{\mu_r^2}{m_b^2}+ \ln\frac{Q_*^2}{\mu_r^2})+3{\hat r}_{2,1} (\ln\frac{\mu_r^2}{m_b^2}+ \ln\frac{Q_*^2}{\mu_r^2})^2+{\hat r}_{1,0} (\ln\frac{\mu_r^2}{m_b^2}+ \ln\frac{Q_*^2}{\mu_r^2})^3]\} a_{s}^{6}(Q_*)+\mathcal{O} [a_{s}^{7}(Q_*)]. ~\label{nonconformal}
\end{eqnarray}
\end{widetext}
Due to its perturbative nature, we expand the solution of $\ln\frac{Q^2_*}{m^2_b}$ as a power series over $a_s(Q_*)$, i.e.
\begin{eqnarray}
\ln\frac{Q^2_*}{m^2_b}=\sum_{i=0}^{2} S_{i}a^i_s(Q_*),  \label{qstar1}
\end{eqnarray}
where $S_i$ are perturbative coefficients that can be determined up to next-to-next-to-leading-log (${\rm N^2LL}$) accuracy by using the known N$^3$LO-level series $\Gamma_3$. By further using the scale displacement relation between the coupling $a_s(Q_*)$ at the $k_{th}$-order and $a_s(m_b)$ as
\begin{displaymath}
a^k_s(Q_*) = a^k_s(m_b) - k \beta_0 \ln \frac{Q_*^2}{m_b^2} a^{k+1}_s(m_b)+\mathcal{O}[a^{k+2}_s(m_b)],
\end{displaymath}
we finally obtain
\begin{eqnarray}
\ln\frac{Q^2_*}{m^2_b}=T_0+T_1 a_s(m_b)+T_2 a^2_s(m_b)+\mathcal{O} [a^3_s(m_b)],
\label{qstar}
\end{eqnarray}
where the coefficients $T_i~(i=0, 1, 2)$ are
\begin{eqnarray}
T_0=&&-\frac{{\hat r}_{2,1}}{{\hat r}_{1,0}}, \\
T_1=&&\frac{2 \beta _0 ({\hat r}_{2,1}^2-{\hat r}_{1,0} {\hat r}_{3,2})}{{\hat r}_{1,0}^2}+\frac{4 ({\hat r}_{2,0} {\hat r}_{2,1}-{\hat r}_{1,0} {\hat r}_{3,1})}{3{\hat r}_{1,0}^2},
\end{eqnarray}
and
\begin{widetext}
\begin{eqnarray}
T_2=&&\frac{5 \beta _1 ({\hat r}_{2,1}^2-{\hat r}_{1,0} {\hat r}_{3,2})}{2 {\hat r}_{1,0}^2}+\frac{16 ({\hat r}_{1,0} {\hat r}_{2,0} {\hat r}_{3,1}-{\hat r}_{2,0}^2 {\hat r}_{2,1})+15 ({\hat r}_{1,0} {\hat r}_{2,1} {\hat r}_{3,0}-{\hat r}_{1,0}^2 {\hat r}_{4,1})}{9 {\hat r}_{1,0}^3} \nonumber \\
&&-\frac{2 \beta _0  (8 {\hat r}_{2,1} {\hat r}_{3,1} {\hat r}_{1,0}-5 {\hat r}_{4,2} {\hat r}_{1,0}^2+4 {\hat r}_{2,0} {\hat r}_{3,2} {\hat r}_{1,0}-7 {\hat r}_{2,0} {\hat r}_{2,1}^2)}{3 {\hat r}_{1,0}^3}+\frac{2 \beta _0^2 (12 {\hat r}_{1,0} {\hat r}_{3,2} {\hat r}_{2,1}-7 {\hat r}_{2,1}^3-5 {\hat r}_{1,0}^2 {\hat r}_{4,3})}{3 {\hat r}_{1,0}^3}.
\end{eqnarray}
\end{widetext}
It is found that $Q_*$ is exactly free of $\mu_r$ at any fixed-order, indicating that the conventional ambiguity of setting $\mu_r$ is eliminated. Such exactly cancellation of $\mu_r$-dependence is due to the fact that, as shown by Eq.(\ref{nonconformal}), the coefficients of $\ln{\mu_r^2}/{m_b^2}$ are exactly the same as those of $\ln{Q_*^2}/{\mu_r^2}$. This shows that one can choose any perturbative value as the renormalization scale to finish the perturbative calculations, and the resultant scale $Q_*$ shall be independent to such choice. Thus, together with the $\mu_r$-independent conformal coefficients, the PMC decay width $\Gamma_3|_{\rm PMC-s}$ shall be independent to the initial choice of the renormalization scale.

As a subtle point, because the N$^3$LL-order and higher-order terms of the perturbative series (\ref{qstar}), e.g. ${\cal O}(a_s^3)$-terms, are unknown, the scale $Q_*$ shall have a residual scale dependence. Such residual scale dependence is different from the arbitrary conventional $\mu_r$-dependence, since it is generally negligible due to a faster pQCD convergence~\cite{Wu:2019mky}. As shall be shown below, the residual scale dependence for a N$^3$LO decay width $\Gamma_{\Upsilon(1S)\to e^+ e^-}$ is negligible due to both $\alpha_s$-suppression and exponential suppression.

To do the numerical calculation, we take the four-loop $\alpha_s$-running behavior, and use $\alpha_s(M_{Z})= 0.1181\pm0.0011$~\cite{Tanabashi:2018oca} to fix the QCD asymptotic scale $\Lambda_{\rm QCD}$. We adopt the fine structure constant $\alpha(2m_b)=1/132.3$~\cite{Jegerlehner:2011mw}. The $b$-quark $\overline{\rm MS}$-mass $\bar{m}_b(\bar{m}_b)= 4.18\pm0.03$ GeV~\cite{Tanabashi:2018oca}, and by using the four-loop relation between the $\overline{\rm MS}$ quark mass and the pole quark mass~\cite{Marquard:2016dcn}, we obtain the $b$-quark pole mass $m_b=4.93\pm0.03$ GeV.

Using Eq.(\ref{qstar}), we obtain
\begin{eqnarray}
{\rm ln}\frac{Q_*^2}{m_b^2} &=& -2.61-72.52 a_s(m_b) + 6089.58 a^2_s(m_b)\nonumber \\
&& \pm |6089.58 a^3_s(\mu)|^{\rm MAX}_{\mu\in[m_b/2, 2m_b]}\nonumber \\
&=&-2.61-1.24+1.77 \pm 0.065, \label{nqstar}
\end{eqnarray}
which leads to $Q_*=1.75\pm0.06$ GeV. Here as an estimation of those contributions from unknown higher-order terms, as suggested by Refs.\cite{Wu:2014iba, Ma:2014oba}, we take the maximum value of $|T_2 a^3_s(\mu)|$ with $\mu \in[m_b/2, 2m_b]$ as a conservative prediction of the magnitude of the uncalculated $a_s^3$-terms, which causes a scale shift $\Delta Q_*=\pm0.06$ GeV. Such a small scale shift ($\sim \pm 3\%$) is reasonable, since the value of $Q_*$ suffers from both $\alpha_s$-suppression and exponential-suppression. Thus, Eq.(\ref{nqstar}) indicates that the typical momentum flow of the decay, $\Upsilon\to e^+ e^-$, is about $1.75$ GeV, which is only half of the usually guessed choice of $3.50$ GeV. Thus by using the present known N$^3$LO pQCD series, the accurate typical momentum flow for $\Upsilon\to e^+ e^-$ can be achieved.

\begin{figure}[htb]
\includegraphics[width=0.450\textwidth]{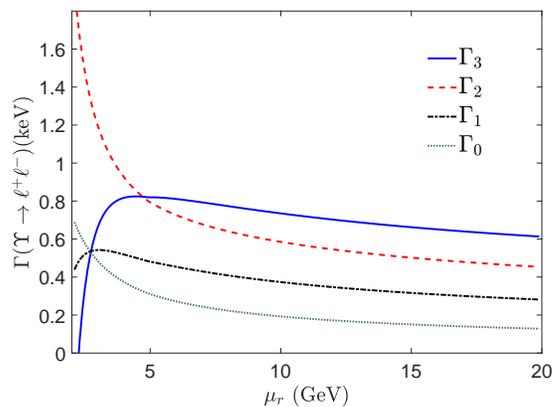}
\caption{The $\Upsilon(1S)$ leptonic decay width $\Gamma_n$ up to $n_{\rm th}$-order QCD corrections as a function of the renormalization scale $\mu_r$ under the conventional sale-setting, where $n=(0,1,2,3)$.}  \label{conv}
\end{figure}

We present the decay width $\Gamma_{n}$ up to $n_{\rm th}$-order QCD corrections under conventional scale-setting in FIG.~\ref{conv}. As expected, if the renormalization scale $\mu_r$ is large enough, e.g. $\mu_r>3$ GeV, the renormalization scale dependence becomes smaller with the increment of loop corrections. On the other hand, it is found that the PMC prediction on $\Gamma_{n}$ under the PMC-s approach is independent to the choice of $\mu_r$ at any fixed $n_{\rm th}$-order.

\begin{figure}[htb]
\includegraphics[width=0.450\textwidth]{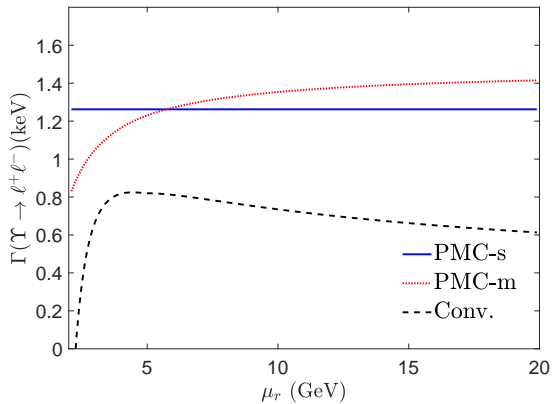}
\caption{The N$^3$LO decay width for the $\Upsilon(1S)$ leptonic decay versus the renormalization scale $\mu_r$ under the PMC-s, PMC-m and conventional scale-setting approaches.}  \label{gamma3}
\end{figure}

To show the scale dependence more explicitly, we present the N$^3$LO decay width $\Gamma_3$ in FIG.~\ref{gamma3}, where the results under conventional and PMC-m scale-setting approaches are presented as a comparison. Firstly, the conventional scale-setting approach leads to the largest renormalization scale dependence, e.g. $\Gamma_{3}|_{\rm Conv.}=[0.665, 0.824]$ keV for $\mu_r\in[3,10]$ GeV, which are only about $50\%-60\%$ of the experimental value $\Gamma^{\rm exp.}\simeq 1.340$ keV. Secondly, such conventional renormalization scale dependence is suppressed by using the PMC-m approach, and a more larger decay width can be achieved. But as has been observed in Ref.\cite{Shen:2015cta}, there is still large residual scale dependence due to a somewhat larger $\mu_r$-dependence for its NLO and NNLO PMC scales, e.g. $\Gamma_{3}|_{\rm PMC-m}=[1.049, 1.353]$ keV for $\mu_r\in[3,10]$ GeV. Such large residual scale dependence for PMC-m approach is reasonable, since the $\Gamma_3$ perturbative series starts at $\alpha_s^3$-order, slight change of its arguments shall result in large scale uncertainty for the decay width. This fact make the process inversely provides a good platform for testing the correct running behavior of the strong coupling constant. Finally, FIG.~\ref{gamma3} shows that, by using the PMC-s approach, the $\Upsilon(1S)$ leptonic decay width is unchanged for any choice of $\mu_r$, e.g. $\Gamma_3|_{\rm PMC-s}\equiv 1.262$ keV.

\begin{table}[htb]
\centering
\begin{tabular}{cccccc}
    \hline
   ~~& ~~$\rm LO$~~& ~~$\rm NLO$~~ & ~~$\rm N^2LO$~~ & ~~$\rm N^3LO$~~& ~~$\rm Total$~~   \\
     \hline
   ~~$\rm PMC-s$& ~~$1.282$~~& ~~$-1.507$~~& ~~$1.583$~~& ~~$-0.096$~~& ~~$1.262$~~ \\
  \hline
  ~~$\rm Conv.$& ~~$0.518$~~& ~~$0.028$~~& ~~$0.491$~~& ~~$-0.258$~~ & ~~$0.779$~~\\
    \hline
\end{tabular}
\caption{Contribution from each order for the N$^3$LO decay width $\Gamma_3$ (in unit: keV) under the PMC-s and conventional ($\mu_r=3.5$ GeV) scale-setting approaches.}  \label{eachorder}
\end{table}

We present the contributions from each order for $\Gamma_3$ in TAB.~\ref{eachorder}. Under conventional scale-setting, the magnitude of the NLO, N$^2$LO and N$^3$LO term is about $5\%$, $95\%$, and $50\%$ of the LO term, respectively. It shows that even at the present known  N$^3$LO level, the conventional pQCD convergence is not as good as required. After applying the PMC, the pQCD convergence is improved, the magnitude of N$^3$LO term is only $\sim 8\%$ of the LO term. More over, as a conservative estimation of the magnitude of the unknown N$^4$LO-terms of the PMC series, we set its value as $\Delta_4=\pm |r_{4,0} a_s^{7}(Q_*)|$. It is negligibly small, e.g. $\Delta_4 \sim \pm 0.002 ~{\rm keV}$.

Moreover, after eliminating the renormalization scale uncertainties via using PMC-s approach, there are still uncertainty sources, such as the $\alpha_s$ fixed-point error $\Delta\alpha_s(M_Z)$, the choices of $b$-quark pole mass $m_b$, the choices of the factorization scale, and etc.

As for the $\alpha_s$ fixed-point error, by using $\Delta\alpha_s(M_Z)=0.0011$~\cite{Tanabashi:2018oca} to fix the $\alpha_s$ value at the required scales, we have $\Lambda_{{\rm QCD},n_f=4}=296\pm16$ MeV, which lead to
\begin{eqnarray}
\Gamma_{\Upsilon({\rm 1S})\rightarrow\ell^+\ell^-}|_{\rm PMC-s} =1.262^{+0.161}_{-0.138} ~{\rm keV} \label{alphaserr}
\end{eqnarray}
and
\begin{eqnarray}\label{alphasconverr}
\Gamma_{\Upsilon({\rm 1S})\rightarrow\ell^+\ell^-}|_{\rm Conv.} =0.779^{+0.054}_{-0.050} ~{\rm keV}.
\end{eqnarray}
As shall show below, such fixed-point error $\Delta\alpha_s(M_Z)$ dominates the error for $\Upsilon(1S)$ leptonic decay width. This indicates that after applying the PMC-s approach, even if we have achieved a renormalization scale-independent conformal coefficients for each perturbative order and have determined the correct momentum flow of the process (being the argument of $\alpha_s$), we still need an accurate referenced fixed-point value $\alpha_s(M_Z)$ so as to a determine an accurate $\alpha_s$ at any scales and hence to achieve a more accurate pQCD prediction. Here, the conventional error of $\Delta\Gamma|_{\rm Conv.}=\left(^{+0.054}_{-0.050}\right) ~{\rm keV}$ is predicted by fixing $\mu_r=3.5$ GeV. Eqs.(\ref{alphaserr}, \ref{alphasconverr}) shows that the conventional error is smaller than the PMC-s one, this is because that the determined effective scale $Q_*=1.75\;{\rm GeV}$ is smaller than $3.50$ GeV, and then the value of $\alpha_s(Q_*)$ is more sensitive to the variation of $\Lambda_{\rm QCD}$.

\begin{figure}[htb]
\includegraphics[width=0.450\textwidth]{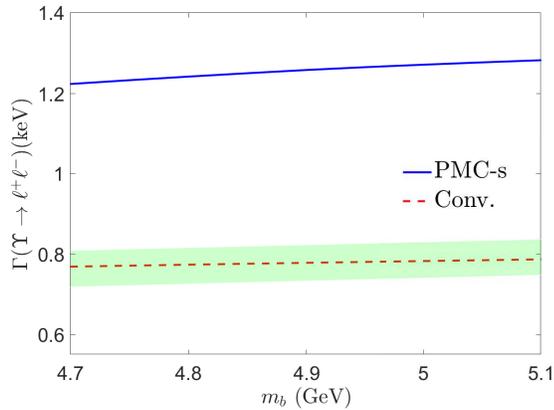}
\caption{The N$^3$LO decay width for $\Upsilon(1S)$ leptonic decay versus $m_b$ under the PMC-s and conventional approaches. The solid line is for PMC-s, which is independent to the choice of $\mu_r$. The error band is for conventional result for $\mu_r\in[3,10]$ GeV, where the dashed line is for $\mu_r=3.50$ GeV, the lower edge is for $\mu_r=10$ GeV and the upper edge is for $\mu_r=4.45$ GeV, respectively. }  \label{changemb}
\end{figure}

The N$^3$LO leptonic decay width $\Gamma_{\Upsilon({\rm 1S})\rightarrow\ell^+\ell^-}$ versus the choice of $b$-quark pole mass $m_b$ is presented in FIG.~\ref{changemb}. If using $m_b=4.93\pm0.03$ GeV, the error $\Delta\Gamma$ shall be negligibly small for both the PMC-s and conventional results under a fixed choice of $\mu_r$.

\begin{table}[htb]
\centering
\begin{tabular}{cccc}
\hline
      & ~~$\mu_h$~~    & ~~$\mu_s$~~    & ~~$\mu_{us}$~~ \\
\hline
      & ~~$+0.057$~~ & ~~$+0.091$~~ & ~~$+0.027$~~ \\
\raisebox {2.0ex}[0pt]{~~$\Delta\Gamma_{3}|_{\mathrm{PMC-s}}({\rm keV})$~~}
      & ~~$-0.066$~~ & ~~$-0.078$~~ & ~~$-0.030$~~ \\
\hline
      & ~~$+0.041$~~ & ~~$+0.053$~~ & ~~$+0.004$~~ \\
\raisebox {2.0ex}[0pt]{~~$\Delta\Gamma_{3}|_{\mathrm{Conv.}}({\rm keV})$~~}
      & ~~$-0.039$~~ & ~~$-0.043$~~ & ~~$-0.005$~~ \\
\hline
\end{tabular}
\caption{The factorization scale errors $\Delta\Gamma_3$ which are calculated by separately varying $\mu_h$, $\mu_s$ and $\mu_{us}$ by $\pm10\%$ of their center values. }  \label{factorscale}
\end{table}

At present, we have no strict way to set the factorization scale of the process, which is usually chosen as the renormalization scale. For the $\Upsilon(1S)$ leptonic decay, the question is much more involved, since it involves three typical factorization scales, i.e. the hard one $\mu_h\sim m_b$, the soft one $\mu_s\sim m_b v_b$, and the ultra-soft one $\mu_{us} \sim m_b v_b^2$, where $v_b \sim \alpha_s(m_b v_b)$~\cite{Bodwin:1994jh} represents the relative velocity between the constituent $b$ and $\bar{b}$ quarks in $\Upsilon$. For definiteness of discussing the factorization scale dependence, we vary the scales $\mu_h$, $\mu_s$ and $\mu_{us}$ within the range of $\pm10\%$ of their center values, and the results are presented in TAB.~\ref{factorscale}. TAB.~\ref{factorscale} indicates that there is still factorization scale uncertainties after applying the PMC-s approach. The conventional factorization scale uncertainties sound relatively smaller, which are due to accidentally cancelation among different terms involving different scales. In fact, if the process involves only one single energy scale, its factorization scale dependence shall be greatly suppressed if we can set the correct momentum flow of the process by applying the PMC, such kind of examples have been found in Top-pair and Higgs boson production processes~\cite{Wang:2014sua, Wang:2016wgw}.

Using the known N$^3$LO terms together with the PMC-s approach, we obtain a more accurate renormalization scale independent prediction
\begin{equation}
\Gamma_{\Upsilon({\rm 1S})\rightarrow\ell^+\ell^-}|_{\rm PMC-s}=1.262^{+0.195}_{-0.175} ~{\rm keV},
\end{equation}
where the errors are squared average of those from $\Delta\alpha_s(M_Z)$, $m_b$, and the choices of the factorization scales. This decay width agrees with the experimental measurement, $\Gamma_{\Upsilon(1S)\to e^+ e^-}|_{\rm Exp.} = 1.340(18)$~keV~\cite{Tanabashi:2018oca}.

\begin{figure}[htb]
\includegraphics[width=0.450\textwidth]{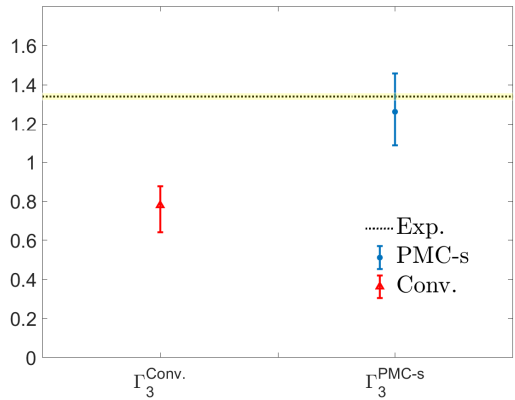}
\caption{The N$^3$LO-level $\Upsilon$ leptonic decay width ($\Gamma_3$) under the PMC-s and conventional scale-setting approaches, respectively. The errors are squared averages of the mentioned theoretical uncertainties. The experimental value~\cite{Tanabashi:2018oca} is also given as a comparison.}  \label{wucha}
\end{figure}

More explicitly, we present our result of the $\Upsilon(1S)$ leptonic decay width up to N$^3$LO level in FIG.~\ref{wucha}, where both the results for PMC-s and conventional scale-setting approaches are presented. FIG.~\ref{wucha} shows that the PMC-s prediction agrees with the experimental measurement within errors, while the conventional prediction is well below the data, e.g.
\begin{equation}
\Gamma_{\Upsilon({\rm 1S})\rightarrow\ell^+\ell^-}|_{\rm Conv.} =0.779^{+0.097}_{-0.137} ~{\rm keV},
\end{equation}
where the errors are squared average of those from $\Delta\alpha_s(M_Z)$, $m_b$, and the choices of the factorization scales and by varying the renormalization scale within the range of $[3, 10]$ GeV.

As a summary, in the paper, we have studied the $\Upsilon(1S)$ leptonic decay width $\Gamma(\Upsilon(1S)\to \ell^+\ell^-)$ by using the PMC-s scale-setting approach. By using the PMC-s approach, we have found that the overall typical momentum flow of the $\Upsilon(1S)$ leptonic decay is $\sim1.75$ GeV, and then a more accurate fixed-order pQCD prediction for $\Gamma(\Upsilon(1S)\to \ell^+\ell^-)$ can be achieved, which agrees with the data and is independent to any choice of renormalization scale. Our present analysis provides another good example for emphasizing the importance of a proper scale-setting approach. Before we draw conclusion of whether there is new physics beyond the standard model for a high-energy process, we need first to get the pQCD prediction as accuracy as possible, especially, we need to set a proper scale, corresponding to the correct momentum flow of the process, for perturbative predictions.

\hspace{0.5cm}

\noindent {\bf Acknowledgement}: This work was supported in part by Natural Science Foundation of China under Grant No.11625520 and No.11847301, and by the Fundamental Research Funds for the Central Universities under Grant No.2019CDJDWL0005.


\begin{thebibliography}{1}

\bibitem{Pineda:1996uk}
  A.~Pineda,
  ``Next-to-leading nonperturbative calculation in heavy quarkonium,''
  Nucl.\ Phys.\ B {\bf 494}, 213 (1997).

\bibitem{Pineda:2001et}
  A.~Pineda,
  ``Next-to-leading log renormalization group running in heavy-quarkonium creation and annihilation,''
  Phys.\ Rev.\ D {\bf 66}, 054022 (2002).

\bibitem{Beneke:1999fe}
  M.~Beneke and A.~Signer,
  ``The Bottom MS-bar quark mass from sum rules at next-to-next-to-leading order,''
  Phys.\ Lett.\ B {\bf 471}, 233 (1999).

\bibitem{Pineda:2006ri}
  A.~Pineda and A.~Signer,
  ``Heavy Quark Pair Production near Threshold with Potential Non-Relativistic QCD,''
  Nucl.\ Phys.\ B {\bf 762}, 67 (2007).

\bibitem{Beneke:2014qea}
  M.~Beneke, Y.~Kiyo, P.~Marquard, A.~Penin, J.~Piclum, D.~Seidel and M.~Steinhauser,
  ``Leptonic decay of the $\Upsilon$(1$S$) meson at third order in QCD,''
  Phys.\ Rev.\ Lett.\  {\bf 112}, 151801 (2014).

\bibitem{Marquard:2014pea}
  P.~Marquard, J.~H.~Piclum, D.~Seidel and M.~Steinhauser,
  ``Three-loop matching of the vector current,''
  Phys.\ Rev.\ D {\bf 89}, 034027 (2014).

\bibitem{Beneke:2005hg}
  M.~Beneke, Y.~Kiyo and K.~Schuller,
  ``Third-order Coulomb corrections to the S-wave Green function, energy levels and wave functions at the origin,''
  Nucl.\ Phys.\ B {\bf 714}, 67 (2005).

\bibitem{Penin:2005eu}
  A.~A.~Penin, V.~A.~Smirnov and M.~Steinhauser,
  ``Heavy quarkonium spectrum and production/annihilation rates to order beta**3(0) alpha**3(s),''
  Nucl.\ Phys.\ B {\bf 716}, 303 (2005).

\bibitem{Beneke:2007pj}
  M.~Beneke, Y.~Kiyo and A.~A.~Penin,
  ``Ultrasoft contribution to quarkonium production and annihilation,''
  Phys.\ Lett.\ B {\bf 653}, 53 (2007).

\bibitem{Beneke:2008cr}
  M.~Beneke and Y.~Kiyo,
  ``Ultrasoft contribution to heavy-quark pair production near threshold,''
  Phys.\ Lett.\ B {\bf 668}, 143 (2008).

\bibitem{Mutuk:2018xay}
  H.~Mutuk,
  ``S-wave Heavy Quarkonium Spectra: Mass, Decays and Transitions,''
  Adv.\ High Energy Phys. {\bf 2018}, 5961031 (2018)

\bibitem{Beneke:2007gj}
  M.~Beneke, Y.~Kiyo and K.~Schuller,
  ``Third-order non-Coulomb correction to the S-wave quarkonium wave functions at the origin,''
  Phys.\ Lett.\ B {\bf 658}, 222 (2008).

\bibitem{Tanabashi:2018oca}
  M.~Tanabashi {\it et al.} [Particle Data Group],
  ``Review of Particle Physics,''
  Phys.\ Rev.\ D {\bf 98}, 030001 (2018).

\bibitem{Wu:2013ei}
  X.~G.~Wu, S.~J.~Brodsky and M.~Mojaza,
  ``The Renormalization Scale-Setting Problem in QCD,''
  Prog.\ Part.\ Nucl.\ Phys.\  {\bf 72}, 44 (2013).

\bibitem{Shen:2015cta}
  J.~M.~Shen, X.~G.~Wu, H.~H.~Ma, H.~Y.~Bi and S.~Q.~Wang,
  ``Renormalization group improved pQCD prediction for ¦´(1S) leptonic decay,''
  JHEP {\bf 1506}, 169 (2015).

\bibitem{Brodsky:2011ta}
  S.~J.~Brodsky and X.~G.~Wu,
  ``Scale Setting Using the Extended Renormalization Group and the Principle of Maximum Conformality: the QCD Coupling Constant at Four Loops,''
  Phys.\ Rev.\ D {\bf 85}, 034038 (2012).

\bibitem{Brodsky:2011ig}
  S.~J.~Brodsky and L.~Di Giustino,
  ``Setting the Renormalization Scale in QCD: The Principle of Maximum Conformality,''
  Phys.\ Rev.\ D {\bf 86}, 085026 (2012).

\bibitem{Mojaza:2012mf}
  M.~Mojaza, S.~J.~Brodsky and X.~G.~Wu,
  ``Systematic All-Orders Method to Eliminate Renormalization-Scale and Scheme Ambiguities in Perturbative QCD,''
  Phys.\ Rev.\ Lett.\  {\bf 110}, 192001 (2013).

\bibitem{Brodsky:2013vpa}
  S.~J.~Brodsky, M.~Mojaza and X.~G.~Wu,
  ``Systematic Scale-Setting to All Orders: The Principle of Maximum Conformality and Commensurate Scale Relations,''
  Phys.\ Rev.\ D {\bf 89}, 014027 (2014).

\bibitem{Shen:2017pdu}
  J.~M.~Shen, X.~G.~Wu, B.~L.~Du and S.~J.~Brodsky,
  ``Novel All-Orders Single-Scale Approach to QCD Renormalization Scale-Setting,''
  Phys.\ Rev.\ D {\bf 95}, 094006 (2017).

\bibitem{Wu:2018cmb}
  X.~G.~Wu, J.~M.~Shen, B.~L.~Du and S.~J.~Brodsky,
  ``Novel demonstration of the renormalization group invariance of the fixed-order predictions using the principle of maximum conformality and the $C$-scheme coupling,''
  Phys.\ Rev.\ D {\bf 97}, 094030 (2018).

\bibitem{Wu:2014iba}
  X.~G.~Wu, Y.~Ma, S.~Q.~Wang, H.~B.~Fu, H.~H.~Ma, S.~J.~Brodsky and M.~Mojaza,
  ``Renormalization Group Invariance and Optimal QCD Renormalization Scale-Setting,''
  Rept.\ Prog.\ Phys.\  {\bf 78}, 126201 (2015)

\bibitem{Ma:2014oba}
  Y.~Ma, X.~G.~Wu, H.~H.~Ma and H.~Y.~Han,
  ``General Properties on Applying the Principle of Minimum Sensitivity to High-order Perturbative QCD Predictions,''
  Phys.\ Rev.\ D {\bf 91}, 034006 (2015)

\bibitem{Wu:2019mky}
  X.~G.~Wu, J.~M.~Shen, B.~L.~Du, X.~D.~Huang, S.~Q.~Wang and S.~J.~Brodsky,
  ``The QCD Renormalization Group Equation and the Elimination of Fixed-Order Scheme-and-Scale Ambiguities Using the Principle of Maximum Conformality,''
  arXiv:1903.12177 [Invited review for Prog. Part. Nucl. Phys. (2019)].

\bibitem{Jegerlehner:2011mw}
  F.~Jegerlehner,
  ``Electroweak effective couplings for future precision experiments,''
  Nuovo Cim.\ C {\bf 034S1}, 31 (2011).

\bibitem{Marquard:2016dcn}
  P.~Marquard, A.~V.~Smirnov, V.~A.~Smirnov, M.~Steinhauser and D.~Wellmann,
  ``$\overline{\rm MS}$-on-shell quark mass relation up to four loops in QCD and a general SU$(N)$ gauge group,''
  Phys.\ Rev.\ D {\bf 94}, 074025 (2016).

\bibitem{Bodwin:1994jh}
  G.~T.~Bodwin, E.~Braaten and G.~P.~Lepage,
  ``Rigorous QCD analysis of inclusive annihilation and production of heavy quarkonium,''
  Phys.\ Rev.\ D {\bf 51}, 1125 (1995).

\bibitem{Wang:2014sua}
  S.~Q.~Wang, X.~G.~Wu, Z.~G.~Si and S.~J.~Brodsky,
  ``Application of the Principle of Maximum Conformality to the Top-Quark Charge Asymmetry at the LHC,''
  Phys.\ Rev.\ D {\bf 90}, 114034 (2014).

\bibitem{Wang:2016wgw}
  S.~Q.~Wang, X.~G.~Wu, S.~J.~Brodsky and M.~Mojaza,
  ``Application of the Principle of Maximum Conformality to the Hadroproduction of the Higgs Boson at the LHC,''
  Phys.\ Rev.\ D {\bf 94}, 053003 (2016).

\end{thebibliography}
\end{document}